# EXTENDED DISTRIBUTED UML-BASED PROTOCOL SYNTHESIS METHOD

Jehad Al Dallal

Department of Information Science, Kuwait University, Kuwait

## ABSTRACT

*Synthesizing specifications for real time applications that involve distributed communication protocol entities from a service specification, which is modeled in the UML state machine with composite states, is a time-consuming and labor-intensive task. Existing synthesis techniques for UML-based service specifications do not account for timing constrains and, therefore, cannot be used in real time applications for which the timing constraints are crucial and must be considered. In this paper, we address the problem of time assignment to the events defined in the service specification modeled in UML state machine. In addition, we show how to extend a technique that automatically synthesizes UML-based protocol specifications from a service specification to consider the timing constraints given in the service specification. The resulting synthesized protocol is guaranteed to conform to the timing constraints given in the service specification.*

## 1. INTRODUCTION

A protocol can be defined as an agreement on the exchange of information between communicating entities. A full protocol definition defines a precise format for valid messages (a syntax), procedure rules for the data exchange (a grammar), and a vocabulary of valid messages that can be exchanged, with the meaning (semantics).

In protocol design, interacting entities are constructed to provide a set of specified services to the service users. While designing a communication protocol, semantic and syntactic errors may exit. Semantic design errors cause the provision of incorrect services to the distributed protocol users. Syntactic design errors cause the protocol to deadlock.

A communication system is most conveniently structured in layers. The service access point (SAP) is the only place where a layer can communicate with its surrounding layers or service users. The layer can have several SAPs. The communication between the layer and its surrounding layers is performed using service primitives (SPs). The SP identifies the type of event and the SAP at which it occurs.

From the user's viewpoint (high level of abstraction), the layer is a black box where only interactions with the user—identified by the SPs—are visible. The specification of the service provided by the layer is defined by the ordering of the visible SPs and by the timing requirements between the SP occurrences. This specification is called service specification (S-SPEC). At a refined level of abstraction, the service provided by the layer is performed using a number of cooperating protocol entities. These protocol entities exchange protocol messages through a communication medium. The protocol specification (P-SPEC) prescribes the exchange of messages between the protocol entities. Figure 1 shows the two abstraction levels of a communication layer. Both S-SPEC and P-SPEC can be modelled using UML state machine.

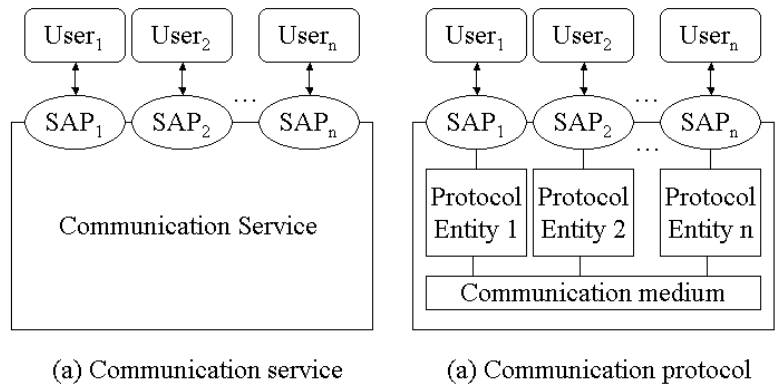

Figure 1. Communication service and protocol concepts

Protocol specifications are much more complex than service specifications because of their refined nature. Therefore, it is quite natural to start the protocol design process from a complete and unambiguous service specification. The construction of a protocol specification from a given service specification is called a protocol synthesis. The synthesis approach is used to construct or complete a partially specified protocol design such that the interactions among the constructed or completed protocol entities proceed without encountering any logical error and, ideally, provide the specified service.

Several protocol synthesis methods for different models have appeared in the literature such as in [1–5]. Only [5] considers the UML state machine model, but it does not consider timing requirements. In this paper, the assignment of the timing constraints to the service specification is discussed. In addition, the UML-based method is extended to synthesize protocol specifications from service specifications containing timing requirements. The resulting protocol specification is proved to conform to the timing constraints provided in the service specification.

The paper is organized as follows: Section 2 gives an overview of the related research. Section 3 discusses the service specification time assignment and introduces the timed protocol synthesis method and a small example. The correctness of the synthesis method is proved in Section 4. Finally, section 5 provides conclusions and discussion of future work.

## 2. RELATED WORK

In this section, an overview of other related research is provided and the basic service-oriented synthesis method introduced in [4] is briefly described.

## 2.1. Other Related Research

The protocol synthesis methods can be classified according to the used models. The used models include finite state machines [2,3], UML state machines [4], Petri-nets [1], and LOTOS-like [5]. Some of the service-oriented protocol synthesis methods consider the timing requirements given in the service specification [3,6], while others do not [1,2,4,5]. The method of dealing with timing constraints provided in the service specifications in [3,6] cannot be directly applied in this paper because a different model is used (i.e., Petri-nets and finite state machines).

UML has been shown to be useful in modelling communication protocols (e.g., [7,8,9]). In [4], a synthesis method that derives protocol specifications from UML-based service specification has been introduced. The derivation process is performed in five steps. In the first step, the S-SPEC is projected onto each SAP to obtain the projected protocol specifications (PR-SPECs). In the second step, a set of transition synthesis rules is applied to the transitions of the PR-SPECs to obtain the primary protocol specifications of the entities (PPE-SPECs). In the third step, $\varepsilon$-transitions and $\varepsilon$-cycles are removed from the PPE-SPECs and a state machine reduction technique is applied to obtain the minimized PPE-SPECs. Then, rules are applied to remodel composite states of multiple regions. Finally, a state machine reduction technique is applied to obtain the minimized PE-SPECs. This method is extended here to consider timing constraints that were ignored in [4].

This paper extends the conference-based version [10] by elaborating more on the synthesis method and providing the correctness proofs for the proposed technique extension.

## 2.2. UML State Machine

The UML state machine [11, 12, 13] is a diagram that consists of states shown in rounded-corner rectangles that are connected with labeled arrows which represent transitions. Each transition can be associated with (1) an event, (2) a set of predicates, and (3) a set of expected actions. To execute a transition, the protocol must be in the accepting state of the transition, the event is executed, and the predicates evaluate to true. The UML syntax for a transition is:

event-name [guard predicate]/action-expression.

A state can be simple, composite, or a submachine. A simple state is a state that does not have any substates. A composite state may contain states of any type. A composite state can include one or more orthogonal regions separated by dashed lines to represent a concurrent behavior. Each region includes substates connected by transitions.

A state can be classified as typical or special. A typical state expresses a stable situation that represents the state context. Special states include initial, final, join, fork, junction, and choice states. An initial state of a state machine represents the starting state of the protocol represented by the state machine. A junction state is used to attach its incoming transitions together. A choice state is used to attach outgoing transitions together. A final state expresses the completion of the protocol that is specified in a composite state region or described by the state machine. The fork state splits an incoming transition into several unlabeled transitions which terminate on states in different regions of a composite state. The join state merges multiple incoming transitions from states of different regions of a composite state into a single, unlabeled outgoing transition.

## 2.3. The Basic Synthesis Method

The synthesis method introduced in [4] uses the UML protocol state machine to model both service and protocol specifications.

To synthesize the protocol specification from the service specification, five steps are followed.
1. Project the service specification S-SPEC onto each SAP to obtain the PR-SPECs. Each PR-SPEC has the same structure as the S-SPEC. Each PR-SPECi has two types of transitions: SP-labeled and unlabeled. The SP-labeled transitions correspond to the S-SPEC transitions assigned to SPs, which are observed at SAPi.
2. Apply transition synthesis rules to the transitions of the PR-SPECs to obtain the primary protocol specifications of the entities (PPE-SPECs). A set of four rules are proposed to cover all possibilities of the different types of source and destination states of any transition. The rules are applied to determine the events associated with the transitions. Transition that are left with no events are associated with $\varepsilon$.
3. Remove $\varepsilon$-cycles and $\varepsilon$-transitions by using algorithms that are described in order to obtain the reduced PPE-SPECs.
4. Remodel all composite states with multiple regions by using two proposed recursive rules to obtain the protocol specifications of the entities (PE-SPECs). The two rules remodel the multi-region composite states of the resulting state machines of the protocol entities to single-region composite states in such a way that all of the possible orderings of the events are preserved and all of the events are executed.
5. Apply a state machine reduction technique to obtain the minimized PE-SPECs.

The resulting protocol entities are proven to be syntactically and semantically free of errors.

## 3. TIMED PROTOCOL SYNTHESIS METHOD

To synthesize timed protocol specifications, the service specification has to be provided with time constraints associated with the S-SPEC transitions. In this section, the time assignment to the S-SPEC transitions is discussed, and the synthesis method for the timed protocol specification is introduced. Finally, a small example is illustrated.

### 3.1. Service specification time assignment

The assignment of the service specification time constraints is performed during the S-SPEC design process. These time constraints are assigned as time intervals associated with the transitions of the UML state machine that models the S-SPEC. The time interval $[min_t, max_t]$ means that the transition $t$ can be executed only within the time T, since the source state of $t$ is visited, where $min_t \le T \le max_t$. The time T includes the waiting time $T_w$, since the source state is visited. If the SP associated with the transition is to be sent from one protocol entity (PE) to another, the time T also includes the time required for sending the SP from the source PE and receiving the SP by the destination PE. The time for sending and receiving an SP from $PE_i$ to $PE_j$ is the delay $d_{ij}$ of the channel between the two PEs. Therefore, $min_t = \min(T_w) + \min(d_{ij})$ and, consequently, $min_t$ has to be greater than or equal to $\min(d_{ij})$. Similarly, $max_t = \max(T_w) + \max(d_{ij})$ and, consequently, $max_t$ has to be greater than or equal to $\max(d_{ij})$. In addition, the $min_t$ and $max_t$ have to be assigned such that $\max(T_w) \ge \min(T_w)$. In other words, $max_t - \max(d_{ij}) \ge min_t - \min(d_{ij})$. Thus, $max_t \ge min_t + (\max(d_{ij}) - \min(d_{ij}))$. In some cases, an SP associated with a transition can be sent to more than one PE (e.g., in Figure 2, $A_1$ is sent to $PE_2$ and $PE_3$). Let X be a set of the protocol entities that can receive the SP. Generally, if an SP associated with a transition $t$ can be sent from $PE_i$ to more than one PE such that each $PE \in X$, the time interval associated with $t$, has to be assigned such that $\forall j \in X$, $min_t \geq min(d_{ij})$ and $max_t \geq max(d_{ij})$. This means that $min_t \geq maximum_{\forall j \in X} (min(d_{ij}))$ and $max_t \geq maximum_{\forall j \in X} (max(d_{ij}))$. Similarly, $\forall j \in X$, $max_t \geq min_t + (max(d_{ij}) - min(d_{ij}))$. This means that $max_t \geq min_t + maximum_{\forall j \in X}(max(d_{ij}) - min(d_{ij}))$.

For example, in Figure 2, the service primitive $A_1$ is sent to $PE_2$ and $PE_3$. You can notice that the conditions $max_t \geq maximum(max(d_{12}), max(d_{13}))$ (i.e., 3>maximum (0.1,0.2)), $min_t \geq maximum(min(d_{12}), min(d_{13}))$ (i.e., 1>maximum(0,0.1)), and $max_t \geq min_t + maximum((max(d_{12}) - min(d_{12})), (max(d_{13}) - min(d_{13})))$ (i.e., 3>1+maximum((0.1-0),(0.2-0.1)) are satisfied.

### 3.2. Synthesis of timed protocol specifications

An automatic synthesis method for the protocol entities from a UML-based service specification is introduced in [4] and summarized in Section 2. In this section, the synthesis method is extended to consider the timing constraints provided in the service specification.

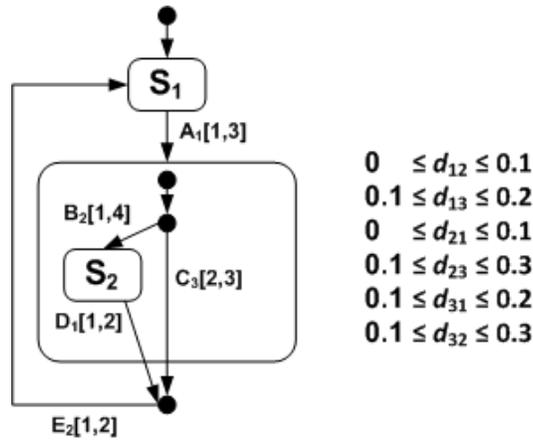

Figure 2. A UML-based service specification example

The first step of the method is extended by associating the transitions of the PR-SPECs with the same time intervals associated with the transitions of the S-SPEC. The PR-SPEC transitions associated with $\varepsilon$-events are not assigned to time intervals.

The second step is extended by applying the modified rules given in Table 1. These modifications consider the timing constraints. The justifications for these extensions are as follows:

Rule 1: In this case, the transition is taking back the service to its initial state and, therefore, a synchronization message is sent to all other PEs. Thus, the channel delays between the $PE_i$ and all other PEs have to be considered. In this case, the maximum and the minimum channel delays among the considered ones are respectively subtracted from $max_t$ and $min_t$ of the transition to obtain the new $max_t$ and $min_t$ values.

Rule 2 and Rules 3 and 4 in the case of $x \neq \emptyset$: In this case, the SP originates from the service user at $SAP_i$. After the occurrence of this SP, other SPs are observed at other SAPs. A synchronization message is sent from $PE_i$ to the other corresponding PEs. Therefore, the channel delays between the $PE_i$ and the other corresponding PEs have to be considered, as illustrated in Rule 1.

Rules 3 and 4 in the case of x = Ø: This rule implies that the flow of control must not be transferred to another protocol entity or service user. Therefore, no channel delays are to be considered. In this case, the same time interval is considered without changing.

The corresponding transitions in PR-SPEC$_x$ and other PR-SPEC (i.e., the last two columns in Table 1) are either associated with $\varepsilon$-events or receiving messages. The transition associated with an $\varepsilon$-event is not assigned a time interval and, therefore, no timing constraints are to be considered. In addition, the transition associated with a receiving message is not assigned a time interval because the time required to execute this transition is part of the channel delay already considered in the above rules.

The rest of the original synthesis method steps remain the same.

Figure 2 shows an S-SPEC example. Figure 3 shows the three PE-SPECs resulting from applying the extended synthesis method. In PE$_1$, the transition associated with the service primitive A has the time interval [1-min(min($d_{12}$),min($d_{13}$)),3-max (max($d_{12}$),max($d_{13}$))] and the transition associated with the service primitive D has the time interval [1-min($d_{12}$),2-max($d_{12}$)]. In PE$_2$, the transition associated with the service primitive E has the time interval [1-min(min($d_{21}$),min($d_{23}$)),2-max(max($d_{21}$), max($d_{23}$))] and the transition associated with the service primitive B has the time interval [1-min($d_{21}$) ,4-max($d_{21}$)]. Finally, In PE$_3$, the transition associated with the service primitive C has the time interval [2-min($d_{32}$),3-max($d_{32}$)].

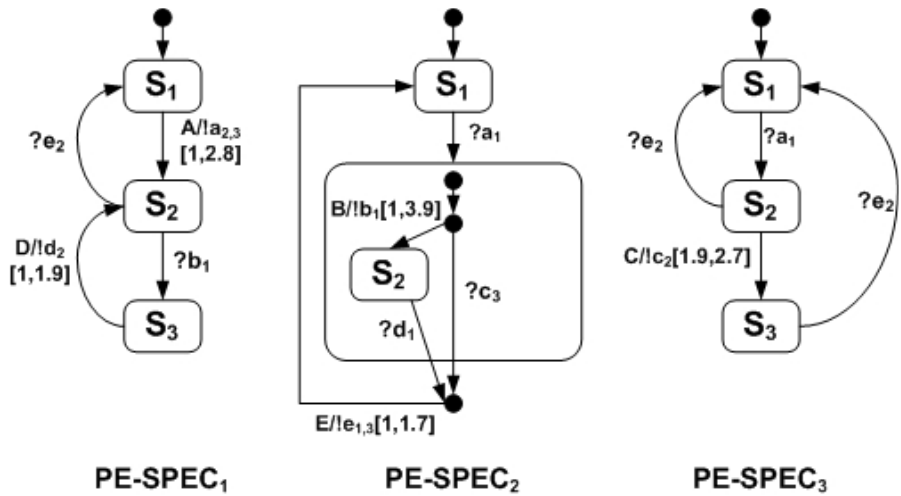

Figure 3. The PE-SPECs obtained by applying the extended synthesis method

Table 1: The extended transition synthesis rules

| Rule ID | Condition ($s_1$ and $s_2$ are source and destination states of the transition $t$ of interest) | x | Event E of transition $t$ in PR-SPEC$_i$ | Corresponding event in PR-SPEC$_x$ | Corresponding event in other PR-SPECs |
|---|---|---|---|---|---|
| 1 | $s_2$ is a stable initial state of the state machine | all SAPs - SAP$_i$ | $E/!e_x$ [$min_t$-min$_{\forall j \in X}$(min($d_{ij}$)), $max_t$-max$_{\forall j \in X}$(max ($d_{ij}$))] | ?$e_i$ | Not applicable |
| 2 | $s_2$ is a final state in the composite state $cs$ Or $s_1$ is a substate of cs and $s_2$ is not a substate of cs Or $s_1$ is a composite state ($s_1=cs$) | $\left[\bigcup_{\forall rg \in cs} OUT(rg)\right] - SAP_i$ | $E/!e_x$ [$min_t$-min$_{\forall j \in X}$(min($d_{ij}$)), $max_t$-max$_{\forall j \in X}$(max ($d_{ij}$))] | ?$e_i$ | Not applicable |
| 3 | $s_2$ is a composite state | $InC(s_2) - SAP_i$ | $E/!e_x$ [$min_t$-min$_{\forall j \in X}$(min($d_{ij}$)), $max_t$-max$_{\forall j \in X}$(max ($d_{ij}$))]  if x≠Ø  E [$min_t$, $max_t$]    if x=Ø | ?$e_i$ | ε |
| 4 | $s_2$ is a simple state | $OUT(s_2) - SAP_i$ | | | |

## 4. PROOF OF CORRECTNESS

Proving the correctness of the synthesis method requires proving that the synthesis method is syntactically and semantically correct. This proof is provided in [4] but without timing constraints. Therefore, to complete the proof, we prove here that the time assignments to the transitions of the PEs as a result of applying the extended synthesis method conform to the time constraints assigned to the transitions of the S-SPEC.

**Lemma 1**. In the PEs, the time T required for executing an SP is $min_{tp} \leq T \leq max_{tp}$ such that the time interval [$min_t, max_t$] is associated with the corresponding transition in the S-SPEC and $min_t \leq min_{tp} \leq T \leq max_{tp} \leq max_t$.

**Proof**: An SP executed in a PE is either (1) not sent to another PE, (2) sent to a service user, or (3) sent to one or more other PEs. In the first two cases, the SP is not sent to another PE and, therefore, no channel delays are to be considered. As a result, in these two cases, the time required to execute the SP in the PE is the same as the time associated with the corresponding transition in the S-SPEC (i.e., ($min_t = min_{tp}$)≤T≤($max_{tp} = max_t$)).

For the third case, the SP is either sent to another PE or sent to more than one other PEs. If the SP is sent to another PE, the time required to execute the SP in the PE is the waiting time since the source state is visited and the channel delay $d_{ij}$. The waiting time is the time associated with the PE transition labeled by SP. This time is T such that $min_t$-min($d_{ij}$)≤T≤$max_t$-max($d_{ij}$). As a result, the time required to execute the SP in the PE (i.e., waiting time + channel delay) is T such that $min_t$-min($d_{ij}$)+$d_{ij}$≤T≤$max_t$-max($d_{ij}$)+$d_{ij}$. This means that $min_{tp}= min_t$-min($d_{ij}$)+$d_{ij}$ and $max_{tp}=max_t$-

max($d_{ij}$)+ $d_{ij}$. Since $min_t$-min($d_{ij}$)+$d_{ij}$ and min($d_{ij}$)≤$d_{ij}$, then $min_t$min($d_{ij}$)+min($d_{ij}$)≤ $min_t$-min($d_{ij}$)+$d_{ij}$≤T. As a result, $min_t$≤$min_{tp}$≤T. Similarly, since T≤$max_t$-max($d_{ij}$)+$d_{ij}$ and $d_{ij}$ ≤ max($d_{ij}$) then T≤$max_t$-max($d_{ij}$)+ $d_{ij}$≤$max_t$-max($d_{ij}$)+max($d_{ij}$). Therefore, T≤$max_{tp}$≤ $max_t$. As a result, in this case, in the PEs, the time T required for executing an SP is T such that $min_t$≤ $min_{tp}$≤T≤$max_{tp}$≤$max_t$.

The last case is when the SP is sent from one PE to more than one other PEs. In this case, the waiting time associated with the transition labeled by SP is T such that $min_t$-minimum$_{\forall j\in X}$(min($d_{ij}$))≤T≤$max_t$ - maximum$_{\forall j\in X}$(max ($d_{ij}$)) where X is the set of the protocol entities that can receive the SP. When considering the channel delays, the minimum time T required to execute the SP is calculated such that $min_t$-minimum$_{\forall j\in X}$(min($d_{ij}$))+ minimum$_{\forall j\in X}$($d_{ij}$)≤T. Since minimum$_{\forall j\in X}$ (min ($d_{ij}$))≤minimum$_{\forall j\in X}$ ($d_{ij}$), then $min_t$-minimum$_{\forall j\in X}$ (min($d_{ij}$))+minimum$_{\forall j\in X}$(min($d_{ij}$))≤ $min_t$-minimum $_{\forall j\in X}$(min($d_{ij}$))+minimum$_{\forall j\in X}$($d_{ij}$)≤T. As a result, $min_t$ ≤$min_{tp}$≤T. Similarly, the maximum time T required to execute the SP is calculated such that T≤ $max_t$-maximum$_{\forall j\in X}$(max($d_{ij}$))+maximum$_{\forall j\in X}$($d_{ij}$). Since maximum$_{\forall j\in X}$ ($d_{ij}$)≤maximum$_{\forall j\in X}$ (max($d_{ij}$)), then T≤ $max_t$-maximum$_{\forall j\in X}$(max($d_{ij}$))+maximum$_{\forall j\in X}$($d_{ij}$) ≤ $max_t$-maximum$_{\forall j\in X}$(max($d_{ij}$))+maximum$_{\forall j\in X}$(max ($d_{ij}$)). Therefore, T≤$max_{tp}$≤$max_t$. As a result, in this final case, in the PEs, the time T required for executing an SP is T such that $min_t$≤ $min_{tp}$≤ T≤$max_{tp}$≤$max_t$.

As a result, for all cases, in the PEs, the time T required for executing an SP is $min_{tp}$≤T≤$max_{tp}$ such that the time interval [$min_t$,$max_t$] is associated with the corresponding transition in the S-SPEC and $min_t$≤ $min_{tp}$≤T≤$max_{tp}$≤$max_t$. ∎

**Lemma 2**. For any sequence of SPs in the S-SPEC executed during the time interval [$min_t$,$max_t$], the corresponding SPs in the PEs are executed within the same or narrowed time interval.
**Proof**: The execution of sequence of *n* SPs in the S-SPEC is performed during the time interval [$min_t$,$max_t$] such that $min_t$ = $min_t$(SP$_1$)+ $min_t$(SP$_2$)+ ….+ $min_t$(SP$_n$) and $max_t$ = $max_t$(SP$_1$)+ $max_t$(SP$_2$)+ ….+ $max_t$(SP$_n$). By Lemma 1, for any SP, $min_t$(SP) in the PE-SPEC is greater than or equal to the $min_t$(SP) in the S-SPEC and $max_t$(SP) in the PE-SPEC is less than or equal to the $max_t$(SP) in the S-SPEC. Therefore, the execution of the sequence of *n* SPs in the PE-SPECs is performed within the same or narrowed time interval [$min_t$,$max_t$]. ∎

**Lemma 3**. The time constraints assigned to the transitions of the PEs as a result of applying the extended synthesis method conform to the time constraints assigned to the transitions of the S-SPEC.

**Proof**: As a result of assigning time intervals to the transitions of the PEs using the extended synthesis method, the execution of any sequence of SPs in the PEs is performed during the same or narrowed time intervals given in the S-SPEC (Lemma 2). Therefore, the time constraints assigned to the transitions of the PEs as a result of applying the extended synthesis method conform to the time constraints assigned to the transitions of the S-SPEC. ∎

## 5. CONCLUSIONS AND FUTURE WORK

In this paper, a synthesis method for protocol specifications from UML-based service specifications is extended such that the timing constraints provided in the service specification are considered in the resulting protocol specifications. This extension makes the synthesis method applicable for real time applications. The extension uses UML state machine for modeling both the service and protocol specifications. In this paper, the assignment of the timing constraints to the service specification is discussed. In addition, the paper shows how to map the timing constraints associated with the transitions of the service specification model to the transitions of the protocol specification models. The maximum and minimum delays of the channels between the protocol entities are considered when mapping the timing constraints in this paper.

The basic synthesis method extended in this paper is limited to the service specifications that have sequential behavior (i.e., only one service primitive can be executed at once). In the future, we plan to extend the basic synthesis method to handle possible concurrent occurrence of service primitives in the service specifications, modeled in the UML state machine by using composite states of multiple regions. In addition, we intend to study the effect of the concurrent behavior of the service specification on the assignment of the time constraints to the service and protocol specifications

**Author**

Jehad Al Dallal received his PhD in Computer Science from the University of Alberta in Canada and was granted the award for best PhD researcher. He is currently working at Kuwait University in the Department of Information Science as an Associate Professor and Department Chairman. Dr. Al Dallal has completed several research projects in the areas of software testing, software metrics, and communication protocols. In addition, he has published more than 80 papers in reputable journals and conference proceedings. Dr. Al Dallal was involved in developing more than 20 software systems. He also served as a technical committee member of several international conferences and as an associate editor for several refereed journals.